\documentclass[letterpaper, 10 pt, conference]{ieeeconf}  

\IEEEoverridecommandlockouts                              

\usepackage{graphics}       
\usepackage{epsfig}         
\usepackage{times}          
\usepackage{amsmath}        
\usepackage{amssymb}        
\usepackage{color}
\usepackage{array}
\usepackage{accents}

\usepackage{stackengine}
\usepackage{url}
\usepackage[colorlinks=true, urlcolor=blue, linkcolor=black]{hyperref}

\title{\LARGE \bf Loop Shaping of Hybrid Motion Control with Contact Transition}

\author{Michael Ruderman$^*$
\thanks{$^*$ University of Agder. Contact: {\tt\small michael.ruderman@uia.no} 
\newline \textcolor[rgb]{0.00,0.07,1.00}{Author's accepted manuscript}
}
}

\begin{document}

\newtheorem{theorem}{Theorem}
\newtheorem{rem}{Remark}
\newtheorem{prop}{Proposition}

\maketitle

\bstctlcite{references:BSTcontrol}

\begin{abstract}
A standard motion control with feedback of the output displacement cannot handle unforeseen contact with environment without penetrating into the soft, i.e. viscoelastic, materials or even damaging the fragile materials. Robotics and mechatronics with tactile and haptic capabilities, and in particular medical robotics for example, place special demands on the advanced motion control systems that should enable the safe and harmless contact transitions. This paper shows how the basic principles of loop shaping can be easily used to handle sufficiently stiff motion control in such a way that it is extended by sensor-free dynamic reconfiguration upon contact with the environment. A thereupon based hybrid control scheme is proposed. A remarkable feature of the developed approach is that no measurement of the contact force is required and the input signal and the measured output displacement are the only quantities used for design and operation.
Experiments on 1-DOF actuator are shown, where the moving tool comes into contact with grapes that are soft and simultaneously penetrable.
\end{abstract}

\section{Introduction}  
\label{sec:1}

Motion control systems that can come into precisely defined or (more importantly) unpredictable contact with objects in the environment have always been the focus of active research, especially in the field of control and robotics, and have done so since the eighties of the last century. For instance, the dynamics stability issues during a contact with stiff environments were recognized and addressed (often in context of industrial robotics), see
\cite{an1987dynamic}, when a manipulator in the force control mode
comes in touch with a stiff and kinematically constrained
environmental object. A milestone was the introduction of the concept of impedance control \cite{Hogan85} and later of hybrid impedance control \cite{anderson1988hybrid}, which enabled a deep understanding of the most important (i.e., physically justifiable) interactions and constraints for the controlled impedance-admittance pair of a mechanical motion system in contact with its environment. Motion control of an unconstrained
manipulation, on the one hand, and force control of a constrained
interaction between the manipulator and its environment, on the
other hand, became quickly to 'stumbling block', especially in
view of controlling \emph{contact transitions}, see e.g.
discussion with experiments in \cite{hyde1994controlling}. For a
former compact survey of the force control of manipulators we refer
to e.g. to the work \cite{yoshikawa2000force}, while a more recent overview of
the force control can be found in the robotic literature like e.g.
Springer Handbook of Robotics \cite{villani2016force}.

The ideas of impedance and admittance control of robotic
manipulators, \cite{Hogan85,anderson1988hybrid}, found quickly a
way and appreciation in motion control for drives and mechatronic
systems \cite{ohnishi1994}. A unified passivity-based control
framework for position, torque and impedance control, which uses
the full state feedback and a dedicated energy shaping with
variable gain strategy was also proposed in
\cite{albu2007unified}. Another focus on decomposing
(correspondingly switching) the control structure led to
\emph{hybrid position/force} controllers, the first one 
(most probably) proposed in \cite{raibert1981}. A widely adopted strategy
of hybrid force/motion control, which aims at controlling the
motion along the unconstrained task directions and the force along the
constrained task directions, is using a certain decomposition
which allows simultaneous control of both the contact force and
end-effector motion in two mutually independent subspaces, see
\cite{villani2016force} for details. Here, a selection strategy for the stiffness/compliance parameters and the desired and feeded back variables must be part of the overall control law and is known to be non-trivial and fundamental for the specification of the control task. Although impedance modulation and reconfiguration (correspondingly switching) are widely used in
the hybrid position/force control in robotics, see e.g.
\cite{ficuciello2015variable}, equally as in other motion control
systems such as hydraulic actuators \cite{pasolli2020hybrid}, the
problems of transition and stability of the structural switching
\cite{liberzon1999} remain among the most relevant. It should be
emphasized that during a contact transition, both a hybrid
position/force control and the process plant itself undergo a
structural change. For sufficiently damped contact transitions and a relatively slow dynamics of an available internal state variable that experiences a threshold value upon contact, a switching strategy based on hysteresis relays can be applied, see e.g.
\cite{ruderman2019switching}. For combining the robustness
property of an impedance control in the stiff contacts with the accuracy
of an admittance control in the soft contacts, various approaches were
proposed in robotics. For instance, a predictive instantaneous
model impedance control scheme was described in
\cite{valency2003}, and a continuous switching (with duty cycle as
design parameter) between the controllers with impedance and
admittance causality was provided (also with experiments) in
\cite{ott2015hybrid}.

An important causality constraint is that no one motion system can
simultaneously impress a force on its environment and impose a
displacement or velocity on it. Only one of both control
variables, either interaction force or relative motion of the
environmental object, can be determined along each degree of
freedom \cite{Hogan85}. This results from an instantaneous power
flow between two or more physical systems. Recall that the power
flow is definable as product of an \emph{effort} and a \emph{flow}
variable, the force and velocity for mechanical systems, respectively.
Given these basic principles, it seems obvious that different
control concepts and an approach for combining them are required
to control the unconstrained and constrained motion and the resulting
contact force.

A frequently appearing question is nevertheless how to handle
those manipulator-environment configurations where the control
design is a-priori constrained by some given structure and/or
application specification and, most importantly, by the available sensors
and their arrangement.

In this work, an intuitively understandable (for standard control design in frequency domain) approach of reshaping the
otherwise stiffly designed feedback controllers is proposed. This
way, a smooth and stable contact transition can be guaranteed
without applying a more complex control structure. Most
importantly, only the measured output displacement is used for
feedback and the designed hybrid motion control does not require
additional force sensors or observers of internal states. The rest
of the paper is organized as follows. In section \ref{sec:2}, we
discuss the types of impedance operators that represent an
interaction with environment and relate them to the control
loop properties required for a contact transition. Section
\ref{sec:3} introduces the thereupon based hybrid motion control
design. A detailed experimental case study with soft but penetrable grape objects coming into contact with the feedback controlled mechanical tool is provided in section \ref{sec:4}. Brief conclusions are in section \ref{sec:5}.

\section{Impedance and shaping of disturbance sensitivity function}
\label{sec:2}

For motion control at large, with one relative degree of freedom
$x$ specified in the generalized coordinates, one can define the
\emph{control stiffness}, cf. \cite{ohnishi1994},
\cite{ruderman2020motion}, as
\begin{equation}\label{eq:2:1}
\kappa = \frac{\partial f(\cdot)}{\partial x(t)} \biggl
|_{t\rightarrow \infty}.
\end{equation}
Here $f(\cdot)$ is the overall net force (in the generalized
coordinates) imposed on the moving body of the motion control
system under consideration. For a matched disturbance force $F(t)$
and a feedback control value $u(t)$, which inherently has
dimension of force in the motion control, the total net force can be
considered as a superposition $f = u + F$. Indeed, when a generic
motion control system
\begin{equation}\label{eq:2:2}
\dot{y} = g(y,u,t)
\end{equation}
with the sufficiently smooth flow map $g(\cdot)$ and the vector of
states variables $y \in \mathbb{R}^n$ (often a vector of relative displacements
and velocities) is in the steady-state, the control should balance the
counteracting disturbance. Let us next fix some controlled
operation point, in terms of $(u_0, x_0)$, and consider this way
reduced \eqref{eq:2:1} in the Laplace domain. Then, one can write
\begin{equation}\label{eq:2:3}
\tilde{\kappa}(s) = \frac{F(s)}{x(s)},
\end{equation}
provided that the Laplace transform of the control stiffness
operator exists. Also we note that an equivalence between $\kappa$
and $\tilde{\kappa}$ is valid only for low frequency range, cf.
\cite{anderson1988hybrid}. Recalling that for a linear
environment, an \emph{impedance} is defined as the ratio of the Laplace
transforms of effort and flow \cite{spong2005}, the mechanical
impedance is (symbolically) written as
\begin{equation}\label{eq:2:4}
Z(s) = \frac{\hbox{force}(s)}{\hbox{velocity}(s)},
\end{equation}
that leads to a generic relationship
\begin{equation}\label{eq:2:5}
\tilde{\kappa}(s) = Z(s) s.
\end{equation}
Since a dynamic interaction between two physical bodies implies
that one must complement the other -- i.e. along any degree for
freedom if one is an impedance then the other must be admittance and
vice versa \cite{Hogan85} -- the condition \eqref{eq:2:5}
becomes crucial. It reveals how the motion control system can be
designed when an environment is classified by $Z(s)$.

\begin{figure}[!h]
\flushleft
(a) \hspace{38mm} (b) \\
\centering
\includegraphics[width=0.95\columnwidth]{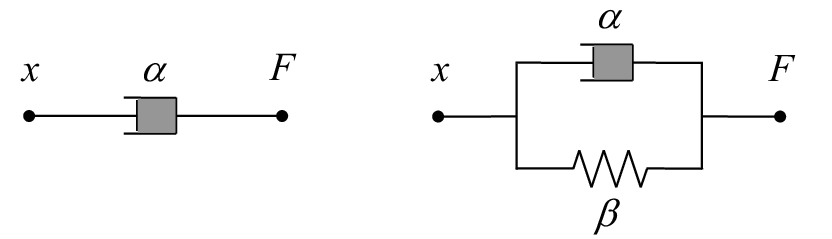}
\caption{Schematic representation of the contact environments: (a)
viscous dashpot, (b) viscoelastic (Kelvin-Voigt type) contact.}
\label{fig:environm}
\end{figure}

Using the mechanical impedance definition \eqref{eq:2:4}, one can
classify the following (typical) contact environments, which are
associated with the corresponding impedance operators. The first
one is the viscous dashpot, shown schematically in Fig.
\ref{fig:environm} (a).
The constitutive equation of the Newtonian fluid in a dashpot,
i.e. $F = \alpha \, \dot{x}$ where $\alpha > 0$ is viscosity,
results in $Z_v(s) = \alpha$, meaning the contact environment is
resistive, cf. \cite{spong2005}. If one expects the environment to
be viscoelastic, i.e. to exhibit also a certain capacitive
behavior, then one can assume a Kelvin-Voigt contact, that is
schematically shown in Fig. \ref{fig:environm} (b). This leads to
the corresponding impedance operator $Z_{ve}(s) = \alpha + \beta /
s$. Note that for modeling the environmental impedance, more
complex structures than those shown in Fig. \ref{fig:environm} can
equally be assumed, and even variable and nonlinear structures
might be considered. This would, however, go far beyond the scope
of this work, and it turns out to be superfluous for a
straightforward design of the linear and hybrid motion controllers.

Now, consider the loop transfer function
\begin{equation}\label{eq:2:8}
L(s) = C(s) G(s)
\end{equation}
of the motion control system with the input-output plant $G(s) =
x(s) (u(s) + F(s))^{-1}$ and feedback controller $C(s)$. The latter 
receives the control error $e(s) = r(s)-x(s)$ as input. Obviously,
the control reference command $r(s)$ conforms to some
application-related specifications before and after a possible
contact with environment and without it. While the
reference-to-output transfer function $H(s) = L(s) (1 +
L(s))^{-1}$ can be determined by shaping $L(s)$, correspondingly
by designing $C(s)$ to be possibly stiff, i.e. having possibly high
bandwidth of $H(s)$ and possibly unity $|H(j\omega)|$ for all
angular frequencies $\omega \in (0, +\infty)$, we focus on the disturbance response to $F(s)$ in the
following. The disturbance-to-output transfer characteristics are given by
\begin{equation}\label{eq:2:9}
S(s) = \frac{x(s)}{F(s)} =  \frac{G(s)}{1 + L(s)}
\end{equation}
and often denoted as \emph{disturbance sensitivity function}, cf.
e.g. \cite{aastrom2021feedback}. While an ideal position (or
velocity) controller should not allow any steady-state or
transient deviations for any force imposition on the mechanical
system, i.e. the controller stiffness should be infinite, cf.
\cite{ohnishi1994}, a hybrid motion controller with contact
transition should allow $S(s)$ to match the $Z(s)$ properties of
the contact with environment. Comparing \eqref{eq:2:3},
\eqref{eq:2:5}, and \eqref{eq:2:9} one can recognize that
\begin{equation}\label{eq:2:10}
S(s) = \tilde{\kappa}^{-1}(s) =  \frac{1}{Z(s)s}.
\end{equation}

To further interpret the results obtained above, it is worth
recalling a fundamental distinction between a mechanical
admittance and impedance \cite{Hogan85}. Multiple physical systems
can be described in one form but not in the other. For instance,
elastic contacts approximated by a non-monotonic constitutive
equation can only be seen as impedance, i.e. $x \mapsto F$, but
not as admittance. Indeed, prior to a mechanical contact is
established, the map $F \mapsto x$ is not given. Similar issue
appears in case of a tangential kinetic friction force, cf.
\cite{ruderman2023analysis}. Nevertheless, the admittance of the
motion control, which is equivalent to disturbance sensitivity
function \eqref{eq:2:9}, can certainly be used at the same moment
as the mechanical contact with environment arises. This will be discussed and used further below in the derivation of hybrid motion control.

\section{Hybrid motion control}
\label{sec:3}

First, we proceed with designing a (sufficiently) stiff feedback
motion control denoted by $C_s$. Following the most simple loop
shaping methodology and assuming the critically damped dominant
pole pair of the closed-loop control system, the reference-to-output
transfer function yields
\begin{equation}\label{eq:3:1}
H(s) = \frac{L(s)}{1+L(s)} = \frac{\omega_0^2}{s^2 + 2 \omega_0 s
+ \omega_0^2}.
\end{equation}
The control specification is then given by only the natural
frequency $\omega_0$ (approximately equal to the bandwidth) of the
closed-loop. Recall that higher $\omega_0$ values imply higher
stiffness of the controlled system. Using the given system
transfer function $G(s)$ and applying the block diagram algebra,
the resulting motion control $C_s(s) = L(s) G^{-1}(s)$ is
\begin{equation}\label{eq:3:2}
C_s(s) = \frac{\omega_0^2}{G (s) s (s + 2\omega_0)}.
\end{equation}
Note that for the system plants $G(s)$ with a relative degree $\leq
2$, the control \eqref{eq:3:2} yields a proper transfer function
and, thus, can be directly implemented.

Now, for designing an impedance controller, assume a sensitivity
function, cf. \eqref{eq:2:10},
\begin{equation}\label{eq:3:3}
S_v(s) = \frac{1}{\alpha s},
\end{equation}
that corresponds to a viscous (dashpot-type) contact with
environment, cf. section \ref{sec:2}. Using \eqref{eq:2:9} and the
block diagram algebra, the resulting impedance controller (denoted further
as viscous) is determined by
\begin{equation}\label{eq:3:4}
C_v(s) = \frac{\alpha s \, G(s)-1}{G(s)}.
\end{equation}
Note that \eqref{eq:3:4} reveals an improper transfer function so
that an additional low-pass filter with a sufficiently high
cut-off frequency needs to be applied in series with $C_v(s)$.

\begin{figure}[!h]
\centering
\includegraphics[width=0.98\columnwidth]{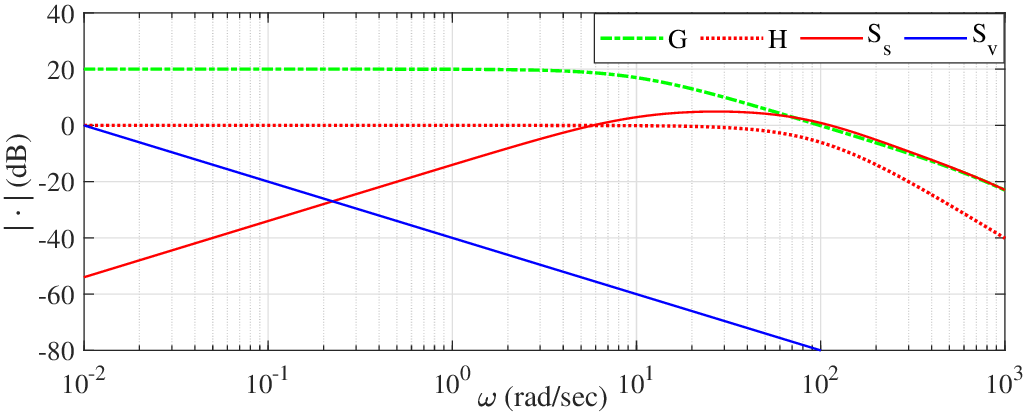}
\caption{Magnitude response of $G(j\omega)$, $H(j\omega)$,
$S_s(j\omega)$, and $S_v(j\omega)$.} \label{fig:FRFanalys}
\end{figure}

The magnitude response of $G(j\omega)$, $H(j\omega)$,
$S_s(j\omega)$, and $S_v(j\omega)$ transfer functions are depicted
in Fig. \ref{fig:FRFanalys} for an exemplary assumed system plant
with two negative real poles at $\{-1000, -10\}$, and the design
parameters $\omega_0 = 100$ and $\alpha = 100$. Note that both
disturbance sensitivity functions are determined according to
\eqref{eq:2:9} for each of the feedback controllers $C_s$ and
$C_v$.
It is easy to interpret from the Bode plot of $S_s(j\omega)$ that
a step-wise disturbance $F(s)$, which appears at the contact
instant, will be largely suppressed by the stiff motion control.
Following to that, the controlled mechanical motion system will continue to move and
penetrate into the environmental object, or moves it away from
itself if the latter is not fixed. Quite the opposite, the
viscous impedance control will react to the step-wise contact
force $F$ by inducing a repulsive relative displacement in the
opposite direction. Here it is worth noting that since it leads to
release from the contact, the disturbance becomes zero and the
relative motion will stop, cf. below with the experiments.

Recall that for achieving stable and smooth contact transitions, a
general strategy of the motion control is to regulate the system
displacement and/or velocity (as conventional manipulators do) and
provide additionally a well-specified disturbance response for
deviations from this motion. According to \cite{Hogan85}, such
disturbance response has the form of an impedance, that may be
then modulated and adapted depending on the control tasks and
environment. Despite such straightforward impedance paradigm, that
gives the name 'impedance control' \cite{Hogan85}, one task that
is not always solvable remains the detection of contact,
correspondingly recognition of the associated deviations from a
well-specified (i.e. nominal) motion. Mostly, the contact forces
are measured by a force sensor connected to the wrist of
manipulator or integrated in the front-end tool. The use of
internal (i.e. not only output displacement) measurements or even
external senors can be found both in the theoretical works, cf.
e.g. \cite{bechlioulis2012} and experimental studies in robotics,
e.g. \cite{DeLuca2006,ficuciello2015variable}.

\begin{figure}[!h]
\centering
\includegraphics[width=0.98\columnwidth]{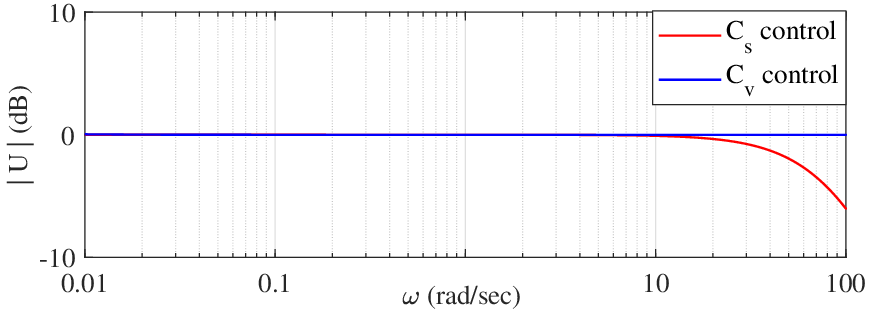}
\caption{Magnitude response of $U(j\omega)$ for $C_s$ and $C_v$
controllers.} \label{fig:ControlSig}
\end{figure}

If the controlled output displacement is the only measurable state
of a motion system (the case that we also consider in this work), the
control reshaping can be triggered exclusively by an information
content of the control signal. Assuming some nominal bound of the
control signal $U$, that is mostly possible for the given nominal
plant $G(s)$, control $C_s(s)$, and reference $r(s)$, an overshot
$|u(t)| > U$ will indicate the appearance of a disturbance force
$F$. Note that this strategy can be used in particular when the
stiff motion controller $C_s$ contains an integral control action.
Indeed, during a compensated steady-state motion, an exceeded
control force is proportional to an additional disturbing force.
Since a dynamic transition from $C_s$ to $C_v$ should not provoke
any undesired transients in direction of the contact with
environment, it is worth examining the
disturbance-to-control-value transfer characteristics which are
given by $U(s) = u(s) / F(s) = C(s)G(s)(1+C(s)G(s))^{-1}$. For
both controllers \eqref{eq:3:2} and \eqref{eq:3:4}, as 
designed exemplary above, this is shown by the magnitude response in Fig.
\ref{fig:ControlSig}.
One can recognize that since $r$ is set to zero for $C_v$, and the
control magnitude response of both $C_s$ and $C_v$ have nearly the
same value, an appearance of the step-wise disturbance $F(t_c)$ at
$t=t_c$ will lead to a decrease of $|u(t_c)| = U$ by the magnitude
equal to $|F|$ for $t > t_c$. Then, the force imposed on the
environmental object under contact drops respectively.

\section{Experimental case study}
\label{sec:4}

\subsection{Motion system and contact scenarios} \label{sec:4:sub:1}

\begin{figure}[!h]
\flushleft
(a) \hspace{38mm} (b) \\[2mm]
\centering
\includegraphics[width=0.45\columnwidth]{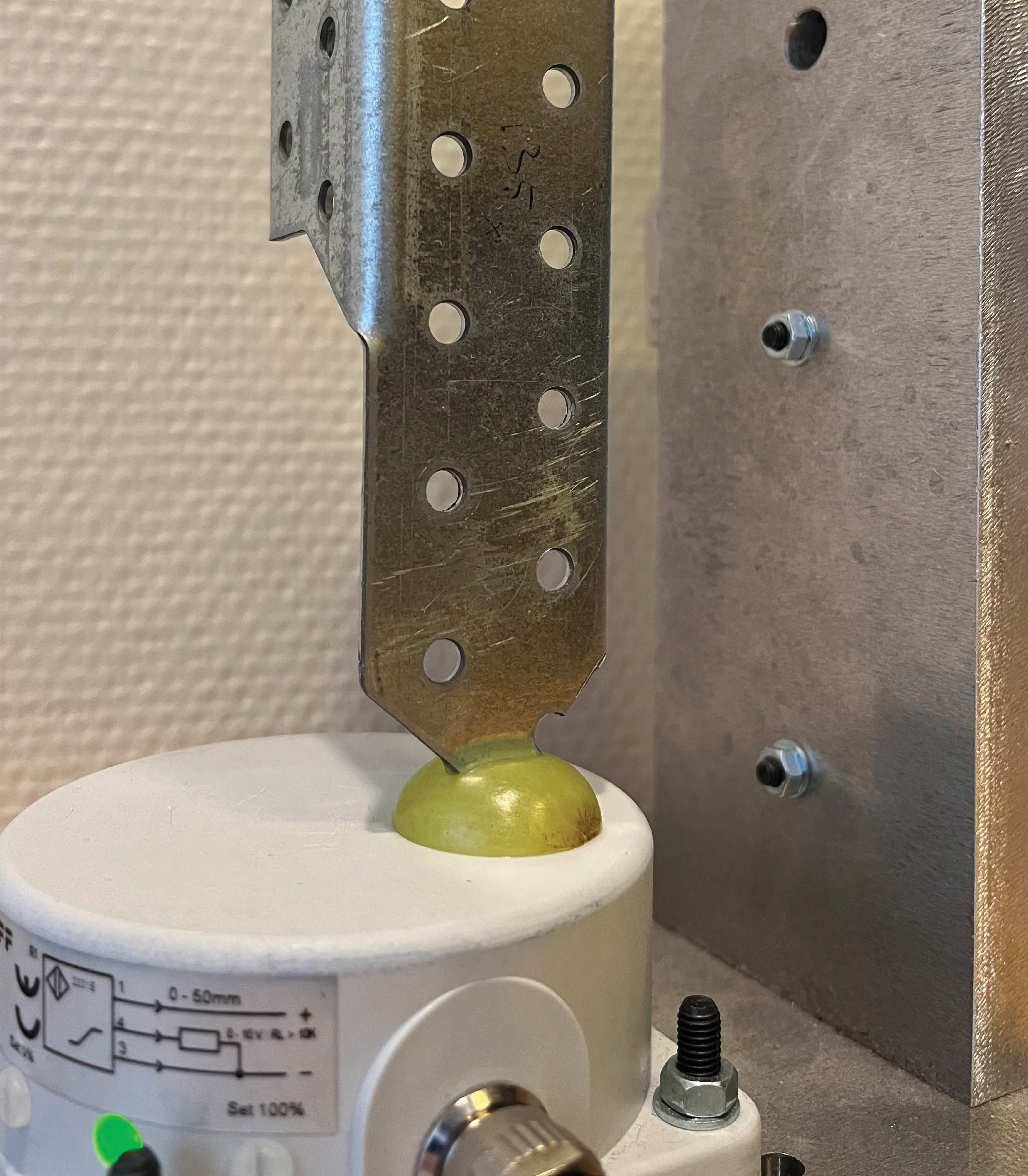} \hspace{3mm}
\includegraphics[width=0.45\columnwidth]{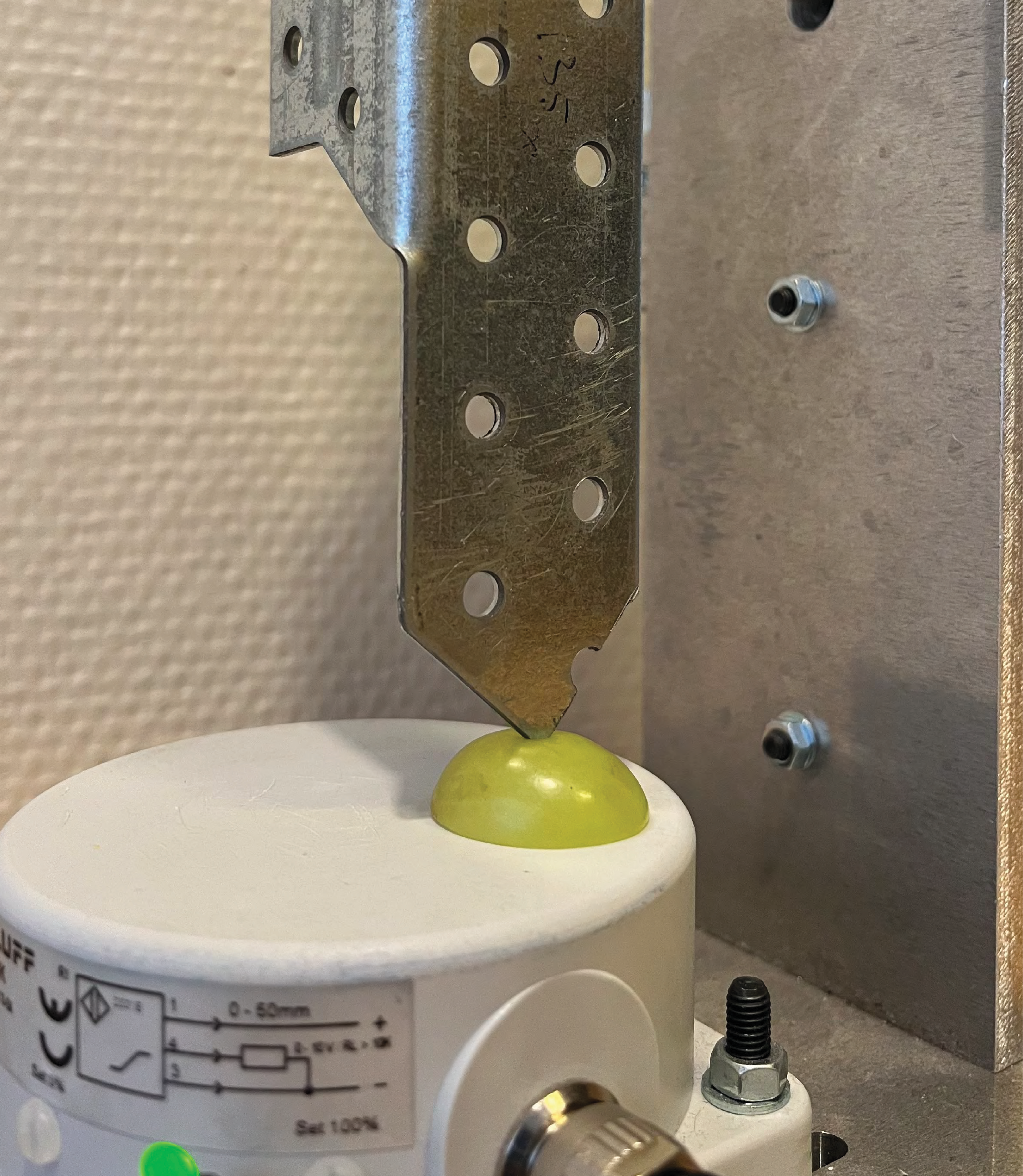}
\caption{Experimental setup of the controlled motion in contact with
a soft environment -- final steady phase after contact with a grape:
(a) stiff motion control $C_s$, (b) hybrid motion control $C_s
\rightarrow C_v$ with reshaped sensitivity.} \label{fig:expsetup}
\end{figure}
In the following, we demonstrate an experimental case study of the
proposed hybrid motion control with a smooth transition when the
moving tool experiences unforeseen contact with environment. The
latter is soft yet penetrable, emulating one of the most critical
applications of the motion control -- robotic assistance for medical
diagnosis and surgery. For that purpose, a half of the grape is
placed in the way of the moving mechanical tool, see Fig.
\ref{fig:expsetup}, which is controlled by using only the
displacement feedback. The relative displacement of the tip of the
tool is sensed remotely, i.e. contactless, cf. Fig.
\ref{fig:expsetup}. Note that the way how the vertical
displacement $x$ is measured provides a relatively high level of
the noise and, thus, represents rather a worst case scenario for
the control application. The system input $v$ (in volt) and output 
$x$ (in meter) are the only quantities
available for the control design and operation. The experimental motion system in actuated by a voice-coil-motor and has one translational degree of freedom. The rigid mechanical
tool is moving in the vertical direction and has a relatively low
displacement range about 0.015 m. The implemented feedback control
is running on the dedicated real-time board with the set sampling
rate of 10 kHz. For more technical details, including the physical
system parameters, an interested reader is referred to
\cite{ruderman2022}.

The nominal system model is given by
\begin{equation}\label{eq:4:1}
x(s) = G(s) v(s) - D = \frac{K}{s(\tau s + 1)} \, v(s) - D,
\end{equation}
where $s$ is the Laplace variable, and $K$ and $\tau$ are the known
system gain and time-constant parameters, respectively. The
constant term $D$ constitutes the nominal disturbance due to the
gravity force which is known. Therefore, the latter is
pre-compensated, so that the system input signal results in
$$
v(t) = D + u(t),
$$
where the feedback controller output $u$ is designed as described above 
in section \ref{sec:3}.

\begin{figure}[!ht]
    \centering
    \includegraphics[width=0.98\columnwidth]{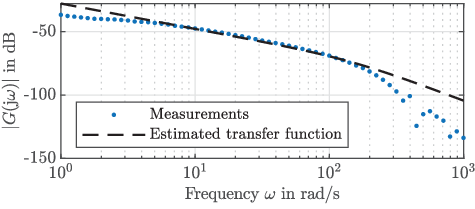}
    \caption{Experimentally measured and least-squares identified magnitude response of the system
    input-output transfer function $G(j\omega)$.}
    \label{fig:IdentFRF}
\end{figure}

Due to a free integrator, cf. \eqref{eq:4:1}, the identification
of the free system parameters was performed in a closed-loop
configuration, see \cite{ruderman2022disturbance} for details. The
least-squares determined parameter values are $K = 0.0408$ and
$\tau = 0.00668$ sec, while the experimentally measured and
identified magnitude response are shown over each other in the Bode
diagram in Fig. \ref{fig:IdentFRF}.
Note that an additional electrical time constant of the
voice-coil-motor dynamics, which is about $0.0012$ sec, is not
explicitly taken into account, equally as not a minor time delay
in the input-output loop of the system. Both are neglected in the
nominal model \eqref{eq:4:1}, cf. \cite{ruderman2022disturbance}.
At the same time, they constitute an additional robustness
criterion for the motion control system under evaluation and, this
way, contribute to a worst-case scenario under evaluation.

\subsection{Evaluated motion control} \label{sec:4:sub:2}

For a sufficiently stiff motion control, i.e. the one without
contact transition and featuring $\bigl| e(s) / F(s) \bigr
|_{s\rightarrow 0} \rightarrow 0$, a standard PID
(proportional-integral-derivative) controller
\begin{equation}\label{eq:4:2}
u_s(s) = C_s(s) e(s) = \bigl(k_p + k_i s^{-1} + k_d s \bigr) \,
e(s),
\end{equation}
is assumed. The controller transfer function is considered as
'stiff' and denoted by $C_s$. The applied robust design procedure,
provided in \cite{ruderman2022disturbance}, rests on an underlying
PD control with a stable pole-zero cancelation (i.e. canceling the
plant time constant $\tau$) and an upper bound of the disturbance
sensitivity function. The determined this way control parameters
are $k_p = 429$, $k_i = 4348$, and $k_d = 2.67$.

Two reshaped 'soft' feedback controllers which allow for a stable
and smooth contact transition are designed in accord with section
\ref{sec:2}. The first one, denoted by $C_v(s)$, is enabling a
purely viscous and well-damped repulsive behavior when contacting
with environment, and it yields
\begin{equation}\label{eq:4:3}
u_v(s) = - \frac{0.0486 \, s^2 + 10.78 \, s}{(0.0272 \, s +
4.08)(\omega_c^{-1} s + 1)} \, x(s).
\end{equation}
Another reshaped feedback controller, denoted by $C_{ve}(s)$, is
more 'stiff' and enables for a viscoelastic behavior when
contacting with environment, cf. section \ref{sec:2}. The
resulting controller has the form
\begin{equation}\label{eq:4:4}
u_{ve}(s) = - \frac{0.0486 \, s^2 + 10.78 \, s}{(0.0272 \, s +
4.08)(\omega_c^{-1} s + 1)} \, x(s) + 429 \, e(s),
\end{equation}
which is similar to \eqref{eq:4:3} but differs from this by including also
a proportional feedback of the control error, cf. with
\eqref{eq:4:2}. The latter makes it possible to press on the
environment with a force that is either proportional to the
control error (in this case the reference value $r(t)$ is to be
considered additionally), or with a constant force equal to $U$.
\begin{figure}[!ht]
    \centering
    \includegraphics[width=0.98\columnwidth]{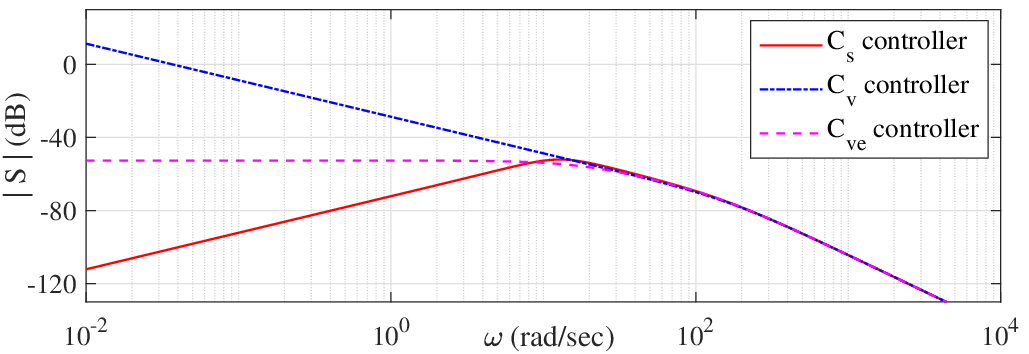}
    \caption{Magnitude response of the disturbance sensitivity function for all
    three designed and evaluated feedback controllers $C_s(s)$, $C_v(s)$, $C_{ve}(s)$.}
    \label{fig:SensTFexper}
\end{figure}
For the second case, which is used in the below experiments,
the proportional control part in \eqref{eq:4:4} is extended by
saturation, i.e. $\mathrm{sat}_U \bigl[k_p \, e(t)\bigr]$. Further
we note that both fractions in \eqref{eq:4:3} and \eqref{eq:4:4}
use a low-pass filter with a sufficiently high cut-off frequency
$\omega_c$; otherwise both control transfer functions are improper
and could not be implemented, cf. section \ref{sec:3}.

The resulted magnitude response of the closed-loop disturbance
sensitivity functions $S(s)$ are compared for all three feedback
controllers (here for $r=0$) in Fig. \ref{fig:SensTFexper}.

\begin{figure}[!ht]
    \centering
    \includegraphics[width=0.98\columnwidth]{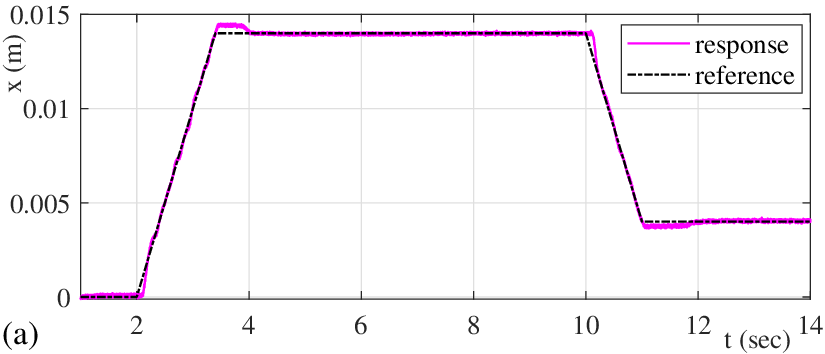}
    \includegraphics[width=0.98\columnwidth]{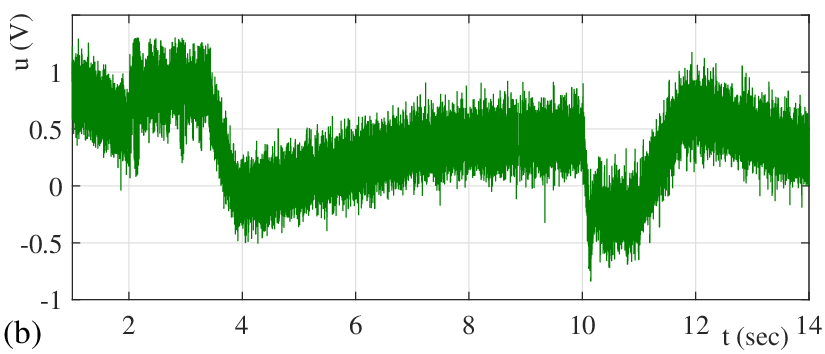}
    \caption{Control response without contacting environment:
    (a) measured displacement $x(t)$ against reference $r(t)$,
    (b) control signal $u(t)$.}
    \label{fig:ExpNoContact}
\end{figure}

The experimentally evaluated motion control scenario is shown in
Fig. \ref{fig:ExpNoContact} (a). The reference trajectory $r(t)$
with one positive and one negative slope, both implying the same
reference velocity magnitude, is tracked by the designed stiff
controller \eqref{eq:4:2}. On the way back, at time $t
> 10$ sec, there is no environmental obstacles and, therefore, no
contact with soft objects, cf. with Fig. \ref{fig:expsetup} where
a soft object was afterwards placed. Note that at the beginning and
especially after the slope segments of trajectory, the control
error $|e(t)|$ increases and takes certain time to settle, cf.
Fig. \ref{fig:ExpNoContact} (a). This is due to nonlinear friction
effects (see \cite{ruderman2023analysis,ruderman2025} for details) which are
not explicitly compensated and need to be mitigated by the
proportional and integral feedback actions only.
The output of the feedback controller $u(t)$ is depicted in Fig.
\ref{fig:ExpNoContact} (b), indicating that it stays in a certain
bound. The control reshaping threshold is set to $U = 1.3$. Recall
that once $|u(t)| > U$, the feedback control changes from its
stiff configuration, i.e. \eqref{eq:4:2}, to a compliant one, i.e.
either \eqref{eq:4:3} or \eqref{eq:4:4} since both controllers $C_v$
and $C_{ve}$ were evaluated. Also recall that the reshaping
threshold value corresponds to the disturbance force which
increases once the stiff motion controller $C_s$ is loading the
contacting object, cf. section \ref{sec:2}.

Following to that, the experimental scenarios with a soft
environmental object, which is a half of the grape placed before
the way back i.e. at time $4 < t < 10$ sec, were evaluated. The
final state is exemplified in Fig. \ref{fig:expsetup}. Three
control configurations were evaluated. The first one is the stiff
control \eqref{eq:4:2} without hybrid reconfiguration to a
compliant controller. The second is the stiff control
\eqref{eq:4:2} which is reconfigured to the viscous control
\eqref{eq:4:3} upon the reshaping threshold $U$. Finally the third
is the stiff control \eqref{eq:4:2} which is reconfigured to the
viscoelastic control \eqref{eq:4:4} upon the reshaping threshold
$U$. Recall that the viscoelastic control \eqref{eq:4:4} is
additionally subject to the saturation at $|u(t)| = U$ since it
contains also the feedback proportional to $e$.
\begin{figure}[!ht]
    \centering
    \includegraphics[width=0.98\columnwidth]{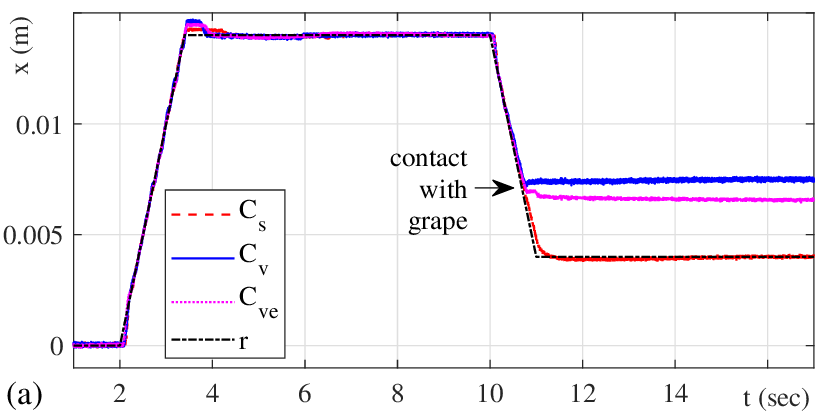}
    \includegraphics[width=0.98\columnwidth]{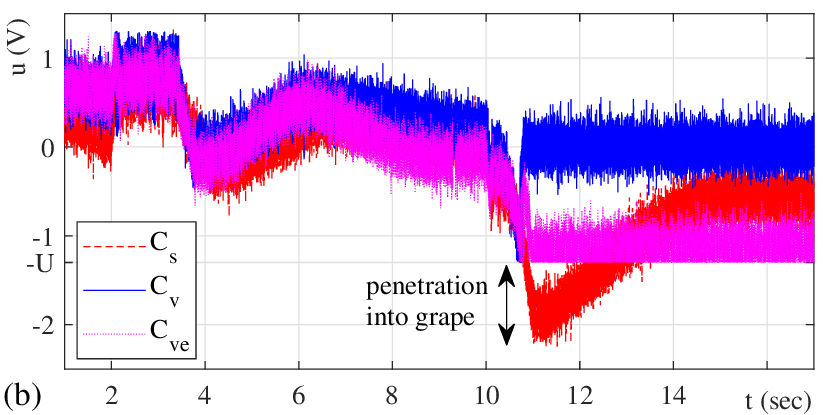}
    \caption{Control response with contacting environment,
    comparing motion controller $C_s$ with hybrid $C_s \rightarrow C_v$
    and $C_s \rightarrow C_{ve}$ reshaping:
    (a) measured displacement $x(t)$ against reference $r(t)$,
    (b) control signal $u(t)$.}
    \label{fig:ExpContact}
\end{figure}
The measured displacement response $x(t)$ and the feedback control
value $u(t)$ are shown in Fig. \ref{fig:ExpContact} (a) and (b),
respectively. One can recognize that the stiff control $C_s$
reaches the back reference position (for $t > 10$ sec), thus
ploughing the mechanical tool into the soft environment, cf. Fig.
\ref{fig:expsetup} (a). Its control value falls below the set
threshold and represents the corresponding force required to
penetrate into the grape. Quite the opposite, the viscous control
$C_v$, activated by exceeding the threshold value, provides a
slightly repulsive response which can be associated with certain
elasticity of the grape surface, cf. Fig. \ref{fig:expsetup} (b).
The controller $C_v$ maintains the contact position while the
control value has a zero mean and almost the same high frequency
pattern as the $C_s$ control has; this is due to the sensor noise and
the corresponding output derivative. A slightly differing behavior
can be seen in case of the viscoelastic controller $C_{ve}$. Due
to a sufficiently large control error $e(t)$ for $t > 11$ sec and,
at the same time, the used control saturation in $C_{ve}$, the
motion system does not penetrate into the grape, but presses it
further with the corresponding threshold magnitude.

The illustrative videos of two of the control experiments reported above, i.e. with the 'stiff' (PID) control \eqref{eq:4:2} and the 'soft' viscous control \eqref{eq:4:3}, can be found online at 
\footnotesize
\href{https://drive.google.com/file/d/1M9Y3YcCdRDVJ4cOXlq45WOhGyT5Jv4II}{https://drive.google.com/file/d/1M9Y3YcCdRDVJ4cOXlq45WOhGyT5Jv4II}\newline
\href{https://drive.google.com/file/d/1JrF01yRCI6ZqB6PMSpZ19hkgjczP8zLX}{https://drive.google.com/file/d/1JrF01yRCI6ZqB6PMSpZ19hkgjczP8zLX}
\normalsize
respectively.

\section{Conclusions}
\label{sec:5}

In this communication, we provided an easily interpretable (in frequency domain) analysis and design for dynamically transforming stiff (admittance) motion control into soft (impedance) control upon contact with constrained and deformable environmental objects. Only the measured output displacement is used. A hybrid control scheme was established, and insightful control experiments involving contact with soft but penetrable objects, e.g., grapes, were conducted.


\bibliographystyle{IEEEtran}
\bibliography{references}

\end{document}